\documentstyle{article}

\begin{document}
\title{A cluster-separable Born approximation for the 3D reduction of the
three-fermion Bethe-Salpeter equation.} 
\author{ J. Bijtebier\thanks{Senior Research Associate at the
 Fund for Scientific Research (Belgium).}\\
 Theoretische Natuurkunde, Vrije Universiteit Brussel,\\
 Pleinlaan 2, B1050 Brussel, Belgium.\\ Email: jbijtebi@vub.ac.be}
\maketitle
\begin{abstract}   \noindent We perform a 3D reduction of the two-fermion
Bethe-Salpeter equation based on Sazdjian's explicitly covariant propagator,
combined with a covariant substitute of the projector on the positive-energy
free states. We use this combination in the two fermions in an external
potential and in the three-fermion problems. The covariance of the two-fermion
propagators insures the covariance of the two-body equations obtained by
switching off the external potential, or by switching off all interactions
between any pair of two fermions and the third one, even if the series giving
the 3D potential is limited to the Born term or more generally truncated. The
covariant substitute of the positive energy projector preserves the equations
against continuum dissolution without breaking the covariance. 
\end{abstract}
 PACS 11.10.Qr \quad Relativistic wave equations. \newline \noindent PACS
11.10.St \quad Bound and unstable states; Bethe-Salpeter equations. \newline
\noindent PACS 12.20.Ds \quad Specific calculations and limits of quantum
electrodynamics.\\\\ Keywords: Bethe-Salpeter equations.  Salpeter's equation.
Breit's equation.\par  Relativistic bound states. Relativistic wave equations.
Three-body problem.\\\\ Short title: A cluster-separable Born approximation...
 \newpage
\tableofcontents

\section{Introduction} The Bethe-Salpeter equation \cite{1,2}  is the usual
tool for the study of relativistic bound states. The principal difficulty in
the treatment of this equation comes from the existence of unphysical relative
time variables. The most usual practice in the two-fermion problem consists in
eliminating the relative time variable (3D reduction). This 3D reduction is
most often based on the replacement of the free Green function by an expression
combining a delta fixing the relative energy and a 3D propagator. The exact
equivalence (in what concerns the physically measurable quantities of the pure
two-fermion problem) with the original Bethe-Salpeter equation can be obtained
by recuperating the difference with the original free Green function in a
series of correction terms to the 3D potential.\par The 3D reduction of the
two-fermion Bethe-Salpeter equation has been performed by many authors [3-19].
All methods are theoretically equivalent at the limit of all correction terms
included. Beyond the two-fermion problem, we have the cases of two fermions in
an external potential ($\,2{1\over2}-$body problem) and of three fermions
(3-body problem). In these cases we meet new difficulties which do not appear
or can be easily solved in the two-fermion case: the Lorentz invariance /
cluster separability problem and the continuum dissolution problem.    
\par --- Lorentz invariance and cluster separability. It is always possible to
render an equation Lorentz invariant by working in the general rest frame
(center of mass reference frame) and by building invariants with the total
4-momentum vector. The real difficulty appears
when we start with a three-fermion equation and "switch off" all interactions
between fermion 3 (for example) and the other ones. We should get a free
equation for fermion 3 and an acceptable equation for the (12) system (cluster
separability). In particular, this implies that the equation for
the (12) cluster must not refer to the global center of mass frame anymore, as the
momentum of fermion 3 enters in the definition of this frame. This equation must
therefore be invariant, explicitly or implicitly (i.e. after rearrangements).\par
--- Solution of the continuum dissolution problem. In the relativistic equations
for several relativistic particles, the physical bound states are degenerate with
a continuum of states combining asymptotically free particles with opposite
energy signs. This often neglected fact forbids the building of normalizable
solutions in the
$\,N\!>\!2$-body problem (including the two-body plus potential problem). The
usual solution consists in including positive-energy projection operators into
the zero-order propagator [20-26]. The modified equations must of course continue
to satisfy the Lorentz invariance / cluster separability requirement.\par 
 In our preceding works on the two-particles in an external potential problem
\cite{22}, and, more recently, on the three-fermion problem \cite{25}, our
approach was based on the cluster separability requirement. In the three-fermion
problem the 3D potential was obtained by adding the three 3D potentials obtained
by the 3D reduction of the three two-fermion equations in the three-fermion rest
frame. Each two-fermion potential depending in general of the total energy of
the subsystem, this two-fermion energy was taken as the three-fermion total
energy, minus the free Dirac hamiltonian of the spectator fermion. The
continuum dissolution was avoided in both cases by choosing a two-fermion
reduction based on a 3D propagator containing a projector on the positive
eigenvalues eigenstates of the Dirac's free hamiltonian. With these
choices the three two-cluster limits are exact: switching off the two interaction
terms with a spectator fermion gives the exact 3D equation which would be
obtained by the reduction of the two-body Bethe-Salpeter equation. From this
exact 3D equation for two fermions it is possible to go back to the original
two-fermion Bethe-Salpeter equation and to perform the 3D reduction again in
another reference frame, so that we can consider this two-fermion equation as
implicitly covariant. This covariance is however quite implicit and global:
neither the approached propagator nor the positive-energy projector being
covariant, the terms of the series giving the 3D propagator are not individually
covariant, so that the implicit covariance becomes an approximated covariance
when the series giving the 3D potential is truncated, for example by keeping only
the Born term.\par In the present work we combine Sazdjian's explicitly covariant
propagator, based on the second-order equations, with a covariant form of the
positive-energy projectors. When used in the 3D reduction of a two-fermion
Bethe-Salpeter equation, this combination leads to a 3D potential given by a
series in which each term is individually covariant. This potential can then be
used in the two-fermion in an external potential problem and in the three-fermion
problem, leading to continuum dissolution-free equations which can be truncated
(for example by keeping only the Born terms) without losing the covariance of
their cluster-separated limits.
\par Although our 3D equations have exact two-cluster limits, they are 
themselves approximations. In the $\,2{1\over2}-$body problem, one should take
into account the modifications brought by the external potential to the fermion
propagators also inside the Bethe-Salpeter irreducible kernel. In the
three-body problem one has to take into account the irreducible three-body
terms at the Bethe-Salpeter equation level and also the three-body terms
generated at the 3D level by the reduction. We tried to do that recently
\cite{25}.
\par 
\hfill\break In section 2 we build a covariant propagator by combining
Sazdjian's covariant propagator with a covariant form of positive-energy
projector. We use it to perform a 3D reduction of the two-fermion
Bethe-Salpeter equation and study the resulting 3D equation (one-body limit,
transition matrix elements, symmetrization of the potential). We present also a
"covariant Salpeter propagator", slightly more complicated but preserving the
particle-antiparticle symmetry. In the two next sections (3 and 4) we exploit
these results by using our 3D potentials in the two fermions in an external
potential problem (with examples) and in the three-fermion problem (with
examples and with a calculation of the two-body limits). Section 5 is devoted
to conclusions.       
        
\section {Two fermions.}

\subsection{Notations.}
\noindent We shall write the Bethe-Salpeter equation for the bound states
 of two fermions \cite{1} as 
\begin{equation}\Phi = G^0 K \Phi,    \label{1}\end{equation}   where
$\Phi$ is the Bethe-Salpeter  amplitude, function of the positions
$x_1,x_2$ or of the momenta 
$p_1,p_2$ of the fermions, according to the representation chosen. The operator
$K$  is the Bethe-Salpeter kernel, given 
by the sum of the irreducible two-fermion Feynman graphs. The operator $G^0$ is
the free propagator, given by the product
$G^0_1G^0_2$ of the two individual fermion propagators
\begin{equation} G^0_i = {1 \over p_{i0}-h_i+i\epsilon h_i}\,\beta_i =
{p_{i0}+h_i\over p_i^2-m_i^2+i\epsilon}\,\beta_i 
\label{2}\end{equation}  where the $h_i$ are the Dirac free hamiltonians
\begin{equation}h_i = \vec \alpha_i\, . \vec p_i + \beta_i\, m_i\qquad (i=1,2). 
\label{3}\end{equation}   We shall define the total (or external, CM, global)
and relative (or internal) variables:
\begin{equation}X = {1 \over 2} (x_1 + x_2)\ , \qquad P = p_1 + p_2\ , 
\label{4}\end{equation} 
\begin{equation}x = x_1 - x_2\ , \qquad p = {1 \over 2} (p_1 - p_2).
\label{5}\end{equation}  and give a name to the corresponding combinations of
the free hamiltonians:
\begin{equation}S = h_1 + h_2\ , \quad s = {1 \over 2} (h_1 -
h_2).\label{6}\end{equation}  We know that, at the no-interaction limit, we
shall have to get a pair of free Dirac equations:
\begin{equation} (p_{10}-h_1)\Psi=0, \qquad (p_{20}-h_2)\Psi=0, 
\label{7}\end{equation}  where $\,\Psi \,$ depends on 
$\,x_1,x_2.\,$ Let us also write their iterated versions
\begin{equation} (p_{10}^2-E_1^2)\Psi=0, \qquad (p_{20}^2-E_2^2)\Psi=0 
\label{8}\end{equation}  with
\begin{equation} E_i=\sqrt{h_i^2}=(\vec p_i^2+m_i^2)^{1\over 2}.  
\label{9}\end{equation}  Interesting combinations can be obtained from the sum
and differences of the equations (\ref{7}) or of the iterated equations
(\ref{8}):
\begin{equation}(P_0-S)\Psi=0, \qquad (p_0-s)\Psi=0, 
\label{10}\end{equation} 
\begin{equation}H^0\Psi=0, \qquad (p_0-\mu)\Psi=0
\label{11}\end{equation}  with
\begin{equation}H^0=2[(p_1^2-m_1^2)+(p_2^2-m_2^2)]_{p_0=\mu}\,\,=\,
\,P_0^2-2(E_1^2+E_2^2)+4\mu^2, \label{12}\end{equation} 
\begin{equation}\mu ={1\over 2P_0}(E_1^2-E_2^2)={1\over
2P_0}(h_1^2-h_2^2)={sS\over P_0}. 
\label{13}\end{equation}

\subsection{Sazdjian's explicitly covariant propagator.}  The free propagator
$\,G^0\,$ can be written 
\begin{equation}G^0=G^{02}\,(p_{10}+h_1)(p_{20}+h_2)
\beta_1\beta_2,\label{2.231}\end{equation} where $\,G^{02}\,$ is the
second-order (spinless) free propagator:  
\begin{equation}G^{02}={1\over(p_{10}^2-E_1^2+i\epsilon)(p_{20}^2-E_2^2
+i\epsilon)}\,.
\label{2.231b}\end{equation} In a two-boson Bethe-Salpeter equation the free
propagator $\,G^{02}\,$ could be approached with the product $\,G^{S2}\,$ of a
constraint $\,\delta(p_0\!-\!\mu)\,$ fixing the relative energy and forcing a
solution of the free difference equation (second equation (\ref{11})) by a 3D
propagator which could be the integral of  $\,G^{02}\,$ with respect to the
relative energy
$\,p_0:$    
\begin{equation}G^{S2}\,=\,\,\delta(p_0\!-\!\mu)\,\int\!
dp_0\,\,G^{02}(p_0)\,=\,-4i\pi\,\delta(p_0\!-\!\mu)\,{\sigma\over
P_0H^0}\label{2.232}\end{equation} with 
\begin{equation}\sigma={1\over 2}(\sigma_1+\sigma_2),\qquad
\sigma_1={P_0+2\mu\over 2E_1},\qquad \sigma_2={P_0-2\mu\over
2E_2}.\label{2.233}\end{equation} On the mass shell
$\,P_0=h_1\!+\!h_2,$ the operator
$\,\sigma_i\,$ is the sign $\,h_i/E_i\,$ of the energy of the fermion i. The
presence of $\,\sigma\,$ annihilates thus the residues of the poles at
$\,P_0=\pm(E_1\!-\!E_2),$ so that $\,G^{S2}\,$ can be written 
\begin{equation}G^{S2}\,=\,-2i\pi\,\,\delta(p_0\!-\!\mu)\,\,{E_1+E_2\over
2E_1E_2}\,\, {1\over P_0^2-(E_1+E_2)^2+i\epsilon}\,.\label{2.234}\end{equation}
The propagator (\ref{2.232}) could easily be written in covariant form if it
did not contain the operator $\,\sigma.\,$  If we replace $\,\sigma\,$ by
$\,\epsilon(P_0)\,$  and add the spinor part of (\ref{2.231}),  we obtain
Sazdjian's propagator
\cite{12} :  
 \begin{equation}G^{SZ}\,=\,{-4i\pi\over \vert
P_0\vert\,H^0}\,\delta(p_0\!-\!\mu)\,(p_{10}+h_1)(p_{20}+h_2)
\beta_1\beta_2\,\label{2.24}\end{equation}   which can also be written in
covariant form
\begin{equation}G^{SZ}\,=\,-2i\pi\,\,\,\delta\,(P\cdot
p-{m_1^2-m_2^2\over2}\,)\,{(p_1\cdot
\gamma_1+m_1)\,(p_2\cdot\gamma_2+m_2)\over
p_1^2+p_2^2-(m_1^2+m_2^2)+i\epsilon}\label{21.2}\end{equation}  or in a form
which exhibits the pole in $\,P_0\!-\!S:$ 
\begin{equation}G^{SZ}\,=\,{-2i\pi\over P_0-S+i\epsilon P_0}\,{P_0+S\over2
\vert P_0\vert}
\,\delta(p_0\!-\!\mu)\,\,\beta_1\beta_2.\label{21.4}\end{equation}

\subsection{Covariant positive-energy projectors.} The combination of two
fermions with opposite energy signs leads to continuum dissolution when an
external potential or a third particle is added. The operator $\,\sigma\,$
would have kill the unwanted poles of Sazdjian's propagator, but we could not
re-introduce it without losing the covariance. We have thus to search for an
invariant substitute of $\,\sigma.\,$  A Lorentz invariant operator forcing a
positive energy for fermion i could be
\begin{equation}\theta_i\,=\,\theta(p_{i0})\theta(p_i^2)=\theta(p_{i0}-\vert\vec
p_i\vert)\label{21.13}\end{equation} as the sign of the energy is invariant for
positive squared 4-momenta.    
 Taking the product of two such operators (one for each fermion) and replacing
$\,p_0\,$ by
$\,\mu\,$ gives
$$\hat\theta\,=\,
\theta\left({P_0\over2}\,+\,{\vec p_1^{\,2}-\vec
p_2^{\,2}\over2P_0}\,-\,\left\vert\,\vec p_1\right\vert
\,+\,{m_1^2-m_2^2\over2P_0}\right)\qquad$$
\begin{equation}
\qquad\theta\left({P_0\over2}\,+\,{\vec p_2^{\,2}-\vec
p_1^{\,2}\over2P_0}\,-\,\left\vert\,\vec p_2\right\vert
\,+\,{m_2^2-m_1^2\over2P_0}\right).\label{4.2.1}\end{equation} Let us assume
that the heaviest fermion is fermion 2. We can then write (\ref{4.2.1}) as   
\begin{equation}\hat\theta\,=\,
\theta\left({P_0\over2}\,+\,{\vec p_1^{\,2}-q_2^2\over2P_0}\,-\,\left\vert \,\vec
p_1\right\vert\right)
\theta\left({P_0\over2}\,+\,{q_2^2-\vec p_1^{\,2}\over2P_0}\,-\,\left\vert
\,\vec p_2\right\vert\right)\label{4.2}\end{equation} with
\begin{equation}q_2\,=\,\sqrt{\,\vec p_2^{\,2}+m_2^2-m_1^2
}\,.\label{4.2.b}\end{equation} The sum of the arguments of the two
$\,\theta\,$ is $\,P_0\!-\!\vert \,\vec p_1\vert\!-\!\vert \,\vec p_2\vert.\,$
The argument of the first $\,\theta\,$ is  $(\,P_0\!-\!\vert \,\vec
p_1\vert\!+\!q_2) (\,P_0\!-\!\vert \,\vec p_1\vert\!-\!q_2)/2P_0,\,$ which
implies that
$\,P_0\,$ must be outside the interval 
$(\vert \,\vec p_1\vert\!-\!q_2,\vert \,\vec p_1\vert\!+\!q_2).\,$ The
combination of these two results implies
$\,P_0\!>\!\vert \,\vec p_1\vert\!+\!q_2.\,$ The second $\,\theta\,$ brings no
supplementary restriction, its argument being always positive when
$\,P_0\!>\!\vert \,\vec p_1\vert\!+\!q_2.\,$ We can thus write
\begin{equation}\hat\theta\,=\,\theta\,(\,P_0-\vert\,\vec p_1\vert-\sqrt{\,\vec
p_2^{\,2}+m_2^2-m_1^2}\,\,).\label{4.2.c}\end{equation} We see that our
projection operator introduces a cutoff on the high $\,\vert\,\vec p_i\vert\,$
for a given
$\,P_0.\,$\par It is interesting to write $\,\hat\theta\,$ also in terms of the
$\,h_i,\,$ for comparison with the
$\,\Lambda^{++}\!\equiv\!\theta(h_1)\theta(h_2)\,$ projector widely used by us
\cite{22,25} and by others, and also for future use in the two-fermion plus
potential problem. We get
\begin{equation}\theta_i\,=\,\theta(p_{i0})\theta(p_i^2)\,=\,\theta(p_{i0})\,
\theta(p_{i0}^2-h_i^2+m_i^2)\label{21.13b}\end{equation}      so that
\begin{equation}\hat\theta\,=\,\left[\theta(p_{10})\,
\theta(p_{10}^2-h_1^2+m_1^2)\,\theta(p_{20})\,
\theta(p_{20}^2-h_2^2+m_2^2)\right]_{p_0=\mu}\,.\label{21.13c}\end{equation}
The $\,p_0\!=\!\mu\,$   constraint gives
\begin{equation}p_{10}\,=\,{P_0\over2}+{h_1^2-h_2^2\over2P_0},\qquad
p_{20}\,=\,{P_0\over2}+{h_2^2-h_1^2\over2P_0}\label{21.13d}\end{equation} and
also
$\,p_{10}^2\!-\!h_1^2=p_{20}^2\!-\!h_2^2\,$ so that the last $\,\theta\,$ which
concerns the heaviest fermion can be omitted. For a given bound state energy
$\,P_0,\,\,\hat\theta\,$ eliminates the mixed energy-sign continuum which
would share this energy. When
$\,P_0\!=\!S,\,$ (\ref{21.13c}) reduces indeed to
$\,\theta(h_1)\theta(h_2).\,$  In fig. 1, we draw a map of the
$\,\hat\theta\!=\!1\,$ region in the
$\,(h_1,h_2)\,$ plane, for  $\,P_0\,$ fixed to 2 (in arbitrary units) and
different values of the lowest mass $\,m_1:\,$ 0, 0.5, 1 (solid line), 1.5, 2,
2.5 and 3 (while $\,m_2\,$ can take any value larger or equal to $\,m_1$). We
draw only the first quadrant, the other ones being the mirror images of the
first one.  The
$\,\hat\theta\!=\!1\,$ region is bounded by the two $\,p_{i0}\!=\!0\,$ curves,
independent of the masses, and by the 
$\,p_1^2\!=\!0\,$ curve corresponding to the lowest mass $\,m_1.\,$ In each
quadrant, this region is divided into four parts, corresponding to the signs of
$\,h_1^2\!-\!m_1^2\,$ and
$\,h_2^2\!-\!m_2^2\,$  (as an exemple, we indicate this partition in the
$\,m_1\!=\!1, m_2\!=\!1.1\,$ case, for which $\,P_0\,$ is thus 0.1 below the
threshold 2.1). When no external potential is present, both expressions are
positive definite, the negative values being for the bound states of the
corresponding fermion in an external potential.\par     While the widely used
$\,\Lambda^{++}\!\equiv\!\theta(h_1)\theta(h_2)\,$ noncovariant projector
simply selects the first quadrant, the
$\,\hat\theta\!=\!1\,$ region is divided into four symmetric parts, one for
each quadrant, but there is a cutoff on the high-$\vert h_i\vert\,$ values. For
moderately relativistic systems, the important regions are near the
$\,(m_1,m_2)\,$ point. \par
 We shall thus finally approach the free propagator by the product of the
covariant Sazdjian propagator with our covariant projector:
\begin{equation}G^{\delta}\,=\,\hat\theta\,G^{SZ}.\label{4.3}\end{equation} 

\subsection{3D reduction of the two-fermion Bethe-Salpeter equation.} 
 We shall write the free propagator as the sum of the zero-order propagator,
plus a remainder:
\begin{equation}G^0=  G^{\delta}+G^R.  \label{23}\end{equation}  The
Bethe-Salpeter equation  becomes then the inhomogeneous equation
\begin{equation}\Phi=G^0K\Phi=(G^\delta +G^R)K\Phi=\Psi +G^RK\Phi,
\label{24}\end{equation}  with
\begin{equation}\Psi=G^\delta K\Phi \qquad (=G^\delta (G^0)^{-1}\Phi).
\label{25}\end{equation}  Solving (formally) the inhomogeneous equation
(\ref{24})  with respect to $\,\Phi\,$ and putting the result into (\ref{25}),
we get
\begin{equation}\Psi=G^\delta K(1-G^RK)^{-1}\Psi=G^\delta K^T\Psi  
\label{26}\end{equation}  where
\begin{equation}K^T=K(1-G^RK)^{-1}=K+KG^RK+...=(1-KG^R)^{-1}K 
\label{27}\end{equation}  obeys
\begin{equation}K^T=K+KG^RK^T=K+K^TG^RK. \label{28}\end{equation}  The
reduction series (\ref{27}) re-introduces in fact the reducible graphs into the
Bethe-Salpeter kernel, but with
$G^0$ replaced by $G^R$. Equation (\ref{26}) is a 3D equivalent of the
Bethe-Salpeter equation. It depends on the choice of
$G^\delta.$ \par
 The relative energy dependence of eq. (\ref{26}) can be easily eliminated:
\begin{equation}\Psi=\delta(p_0\! -\! \mu)\,\psi \label{29}\end{equation}  and
$\,\psi\,$ satisfies the equation               
\begin{equation}\psi\,=\,{-2i\pi\over P_0-S+i\epsilon P_0}\,
\,{P_0+S\over2\,\vert P_0\vert}\,\hat\theta\,\beta_1\beta_2\,K^T(\mu,\mu)
\,\,\psi\label{4.4}\end{equation} with
\begin{equation}\beta_1\beta_2K^T(\mu,\mu)\,\equiv\,\int dp_0' dp_0
\delta(p'_0\! -\!\mu)\beta_1\beta_2K^T(p_0',p_0)\delta(p_0\!
-\!\mu).\label{32}\end{equation} Note that we write $\,(p'_0,p_0)\,$ but
$\,(\mu,\mu),\,$ as we keep $\,\mu\,$ in operator form. This operator is
diagonal in the spatial momentum space. The eigenvalue will depend on the
position of $\,\mu\,$ in the formula: the eigenvalue of the first $\,\mu\,$ in
(\ref{32}) will be built with the final momenta and that of the last
$\,\mu\,$ will be built with the initial momenta. Defining
\begin{equation}V\,=-2i\pi\,\hat\theta\,\beta_1\beta_2K^T(\mu,\mu)\,\hat\theta
\label{31}\end{equation} we get
\begin{equation}\psi\,=\,{1\over P_0-S+i\epsilon P_0}\,
\,{P_0+S\over2\,P_0}\,V\,\psi.\label{4.4f}\end{equation} We included a final
$\,\hat\theta\,$ in the definition of $\,V,\,$ as the wave function at right is
already projected, and we replaced $\,\vert P_0\vert\,$ by $\,P_0,\,$ which is
always positive in the $\,\hat\theta\!=\!1\,$ region.\par        The inversion
of the reduction is given by
\begin{equation}\Phi\,=\,(\,1-G^RK\,)^{-1}\,\Psi\,=\,
(\,1+G^RK^T\,)\,\Psi\,=\,(\,1+G^0K^T-G^{\,\delta}
K^T\,)\,\Psi\,=\,G^0K^T\,\Psi\label{32a}\end{equation}  or, explicitating the
relative energy variable
\begin{equation}\Phi(p'_0)\,=\,G^0(p'_0)\,K^T(p'_0,\mu)\,\psi.
\label{32b}\end{equation}   

\subsection{Covariant Salpeter equation.} A Bethe-Salpeter equation with an
instantaneous kernel (i.e. a kernel independent of the relative energy) leads
directly to Salpeter's 3D equation \cite{3}:
\begin{equation}\psi\,=\,{1\over P_0-S+i\epsilon
P_0}\,V^S\,\psi\label{s1}\end{equation} with
\begin{equation}V^S\,=\,-2i\pi\,\tau\,\beta_1\beta_2\,K\,\tau^2\label{s2}
\end{equation}
\begin{equation}\tau\,=\,\theta(h_1)\,\theta(h_2)-\theta(-h_1)\,\theta(-h_2).
\label{s3}\end{equation}   
When the kernel is not instantaneous, we can make an expansion based on the
approached propagator
\begin{equation}G^{\delta S}\,=\,\delta(p_0\!-\!\mu)\,\int\!
dp_0\,\,G^0(p_0)\,=\,-2i\pi\,\delta(p_0\!-\!\mu)\,{\tau\over P_0-S+i\epsilon
P_0}\,\beta_1\beta_2\label{s4}\end{equation} in the same way as above. This
operator $\,\tau,\,$ which projects on the $\,\tau^2\!=\!1\,$ subspace with a
change of sign for the negative-energy part, is also a continuum
dissolution-preserving operator. \par A
"covariant Salpeter equation" can easily be obtained by combining Sazdjian's
covariant operator with a covariant substitue of $\,\tau\,$ instead of a
covariant substitute of
$\,\theta(h_1)\theta(h_2).\,$ We could indeed use
\begin{equation}\theta^S\,=\,\bigg[\,\big[\theta(p_{10})\,\theta(p_{20})-
\theta(-p_{10})\,\theta(-p_{20})\big]\,\theta(p^2_{1})\,\theta(p^2_{2})
\,\bigg]
_{p_0=\mu}.\label{s5}\end{equation} Our 3D equation would then again be
(\ref{4.4f}), but with the potential
\begin{equation}V\,=-2i\pi\,\theta^S\,\beta_1\beta_2K^{T}(\mu,\mu)\,
(\theta^S)^2
\label{s6}\end{equation} where  $\,K^{T}\,$ must now be built using
$\,\theta^S\,$ instead of $\,\hat\theta.$ \par
As the second term of (\ref{s3}) does not
contribute much in practical calculations, it is often omitted. For the same
reason, we shall omit the second term of (\ref{s5}) and use simply
$\,\hat\theta\,$ in the present work. From a more fundamental point of view,
however, the use of $\,\tau\,$ or of $\,\theta^S\,$  preserves a
particle-antiparticle symmetry which is an important feature of relativistic
theories. 
   
\subsection{One-body limit.} If we go to the center of mass  reference frame,
write
$\,P_0=m_2+W_1\,$  (for $\,P_0>0\,$) and make
$\,m_2\to\infty\,$ in the 3D equation (\ref{4.4f}), we get the equation
\begin{equation}\psi\,=\,{1\over W_1-h_1+i\epsilon}\,
\, V_{\infty}\,\psi,\qquad
V_{\infty}=\lim_{m_2\,\rightarrow\,\infty}\,V.\label{4.8}\end{equation} In
\cite{17}, we showed that a 3D reduction performed with a propagator $\,\Lambda
G^{SZ},\,\Lambda\,$ being a projector, leads to a 3D potential given by
\begin{equation}V^{\Lambda}\,=\,\Lambda\,\left[\,V^{SZ}\,+\,V^{SZ}\,
{1-\Lambda\over
P_0-S}\,V^{SZ}\,+\cdots\right]\,\Lambda.\,\label{4.8b}\end{equation} We also
showed that the limit of $\,V^{SZ}\,$ (let us call it $\,V^C)\,$ is given by
the limit of the Born term (a Coulomb potential in QED). Since the limit of
$\,\hat\theta\,$ is 
$\,\theta(W_1-\vert\,\vec p_1\vert),\,$  we have thus 
\begin{equation}V_{\infty}\,=\,\theta(W_1-\vert\,\vec
p_1\vert)\,\,V^C\left(1\,-\,{\theta(\vert\,\vec p_1\vert-W_1)\over
W_1-h_1}\,\,V^C\right)^{-1}\theta(W_1-\vert\,\vec
p_1\vert).\label{4.9}\end{equation} Equation (\ref{4.8}) is related to the
equation
\begin{equation}\psi^C\,=\,{1\over W_1-h_1+i\epsilon}\,
V^C\,\psi^C\label{4.8.1}\end{equation} which describes the light fermion 1 in
the potential generated by the heavy fermion 2. We have
\begin{equation}\psi\,=\,\theta(W_1-\vert\,\vec
p_1\vert)\,\psi^C.\label{4.10}\end{equation} Equation (\ref{4.8}) is thus the
exact equation for the projection (\ref{4.10}) provided the expansion of
(\ref{4.9}) is not truncated.  If, however, the two-body potential is truncated
to the Born term, its one-body limit will be the projection of $\,V^C\,$
instead of $\,V_{\infty}.$ The introduction of the projector
$\,\hat\theta\,$ has destroyed the exact one-body limit of the Born
approximation, as it would also be the case with other continuum
dissolution-preserving operators like
$\,\Lambda^{++}.$

\subsection{Transition matrix elements.} The 3D off mass shell transition
matrix element is
$$T^{3D}\,=\,V\,+\,V\,{1\over P_0-S+i\epsilon P_0}\,
\,{P_0+S\over2\,P_0}\,V\,+\cdots\label{t1}$$
\begin{equation}=\,-2i\pi\,\hat\theta\,\beta_1\beta_2\,\left[K^T+K^T
G^{\delta}K^T+\cdots
\right](\mu,\mu)\,\hat\theta\label{t1bis}\end{equation}
$$K^T+K^TG^{\delta}K^T+\cdots=K(1\!-\!G^RK)^{-1}
(1-G^{\delta}K(1\!-\!G^RK)^{-1})^{-1}$$
\begin{equation}K(1\!-\!G^RK\!-\!G^{\delta}K)^{-1}=K(1\!-\!G^0K)^{-1}\,=\,T
\label{t2}\end{equation}
so that
\begin{equation}T^{3D}=\,-2i\pi\,\hat\theta\,\beta_1\beta_2\,T(\mu,\mu)\,
\hat\theta.
\label{t3}\end{equation} On the positive-energy mass shell
$\,P_0\!=\!E_1+E_2,\,\, \hat\theta\!=\!1\,$ and
$\,T(\mu,\mu)\,$ is the physical on mass shell scattering matrix element of
field theory. Although the Bethe-Salpeter equation (\ref{1}) was written only
for bound states, our final 3D equation can still be used for describing the
scattering.

\subsection{Symmetric form.}  In our 3D equation (\ref{4.4f}), written in the
form
\begin{equation}(P_0-S)\,\psi\,=\,{P_0+S\over
2P_0}\,V\,\psi\label{y1}\end{equation} the interaction term (the operator
before $\,\psi\,$ at right) is not symmetric and depends on
$\,P_0\,$ via the operator before $\,V\,$ and also via $\,V\,$ itself. This
does not excludes a real energy spectrum, but this can not be seen directly on
(\ref{y1}). In the two-fermion framework, this is not a problem as there exist
many ways of transforming (\ref{y1}) into a symmetric equation. We shall
however use these two-fermion interaction terms in a three-fermion equation
below (in section 4) and, in this new framework, their symmetrization will not
be possible anymore (it would demand a different transformation for each
two-fermion pair). In order to be sure to get a real energy
spectrum from our three-fermion equation, we should thus symmetrize the
interaction terms at the two-fermion level. This can be done by simply
multiplying (\ref{y1}) by $\,1\!+\!V/2P_0,\,$ which gives
\begin{equation}(P_0-S)\,\psi\,=\,U\,\psi,\label{y2}\end{equation}
\begin{equation}U\,=\,V\,-\,\{P_0\!-\!S\,,{V\over2P_0}\}\,+\,{V\over2P_0}
\,(P_0+S)\,
{V\over2P_0}.\label{y3}\end{equation} In a 3D reduction based on the
noncovariant projector (\ref{s4}), we would have simply
$\,V^S.\,$ The first-order energy shift $\,-\!\!<V^2/2P_0>\,$ corresponding to
the difference
$\,U\!-\!V\,$ would then be included in the contribution of the first ladder
term
$\,KG^{RS}K,\,$ which enters in the definition of $\,V^S.$         

\section{Two fermions in an external potential.}

\subsection{3D equation.}

 The two-fermion in an external potential problem is already much more
complicated than the pure two-fermion problem, although it exhibits some 
simplifying features when compared with the three-fermion problem. The
principal new difficulty is the non-conservation of the total spatial momentum
$\,\vec P.\,$ In the pure two-fermion case, this conservation law forbids the
mixing of the physical bound states with the mixed-energy states (continuum
dissolution). When an external potential is present, it becomes possible to get
any given energy in an infinity of ways by combining fermions with opposite
free energy signs. The positive-energy bound states will then not be
normalizable, being mixed with a continuum.
\par The two-fermion plus external potential equations can easily be obtained
with  the simple generalizations \cite{22}
\begin{equation}h_i\,\to\,h_{ie}\,=\,\vec \alpha_i\, . \vec p_i + \beta_i\,
m_i\,+\,V_i(\vec x_i,\gamma_i)\label{5.1}\end{equation}
 where $\,V_i\,$ is the external potential acting on fermion i. These
substitutions must be done in the free Bethe-Salpeter propagator $\,G^0\,$ and
in the manipulations leading to the 3D equation (this is an  approximation, as
we should also modify the fermion propagators inside the irreducible graphs,
but this gives the exact cluster-separated limits). We get
\begin{equation}P_0\,\psi\,=\,\left[\,S_e\,+\,{P_0+S_e\over2\,
\,P_0}\,\,V_e\,\right]\,\psi,\label{4.2.2}\end{equation} with
\begin{equation}S_e\,=\,h_{1e}+h_{2e}\,=\,h_1+h_2+V_i(\vec
x_i,\gamma_i)+V_2(\vec x_2,\gamma_2),\label{p3}\end{equation}
\begin{equation}V_e\,=\,-2i\pi\,\hat\theta_e\,\int dp_0' dp_0
\delta(p'_0\! -\!\mu_e)\beta_1\beta_2K^T_e(p_0',p_0)\delta(p_0\! -\!\mu_e)
\,\hat\theta_e,\label{p4}\end{equation}
\begin{equation}\mu_e\,=\,{1\over2P_0}\,(h^2_{1e}-h^2_{2e})\,
\label{p5}\end{equation}
\begin{equation}\hat\theta_e\,=\,\theta\,({P_0\over2}+\mu_e)\,
\theta\,({P_0\over2}-\mu_e)\,
\theta\,(\,\big[{P_0\over2}+\mu_e\big]^2-h^2_{1e}+m^2_1\,)\,,
\label{p6}\end{equation}
\begin{equation}K_e^T\,=\,K+KG_e^RK+\cdots\label{p7}\end{equation}
$$G^R_e(p_0)=\left[\,{1\over {1\over2}P_0+p_0-h_{1e}+i\epsilon
h_{1e}}\,\,\,{1\over {1\over2}P_0-p_0-h_{2e}+i\epsilon h_{2e}}\right.$$
\begin{equation}+\,\left. {2i\pi\over P_0-S_e+i\epsilon
P_0}\,\,{P_0+S_e\over2P_0}\,\,\delta(p_0\!-\!\mu_e)\,\hat\theta_e\,\right]
\beta_1\beta_2\label{p8}\end{equation}       Besides the free-free continuum,
the
$\,(h_{1e},h_{2e})\,$ spectrum (for which we can use figure 1) contains now
bound-free or free-bound combinations (lines) and bound-bound combinations
(points). The equations are written in the laboratory reference frame, defined
as the reference frame in which the external potential is static. When the
mutual interaction is switched off, we get a pair of independent Dirac
equations in the external potential. When the external potential is switched
off, we get the equation of a system of two mutually interacting fermions. This
last equation can be written in covariant form, the contribution of each term
of (\ref{p7}) being separately invariant. We can keep only the Born term (only
$\,K\,$ in $\,K^T_e\,$ and only the ladder term in $\,K\,$) without losing this
covariance.

\subsection{Examples.}
\subsubsection{Two electrons in the field of a nucleus.} In this case, the
external potentials are given by (in configuration space):
\begin{equation}V_1\,=\,{-Ze^2\over\vert\vec x_1\vert},\qquad
V_2\,=\,{-Ze^2\over\vert\vec x_2\vert}.\label{3ee1}\end{equation} For the
mutual interaction kernel we have to choose a gauge. In Feynman's gauge, the
Born term is
\begin{equation}K(p',p)\,=\,-\,{2ie^2\over(2\pi)^3}\,{1\over
k^2+i\epsilon}\,(\gamma_1\cdot\gamma_2),\qquad k=p=p'.\label{3ee2}\end{equation}
In Coulomb's gauge, it is
\begin{equation}K(p',p)\,=\,-\,{2ie^2\over(2\pi)^3}\,[\,\,{\beta_1\beta_2\over
-\vec k^2}\,-{1\over
k^2+i\epsilon}\,(\vec\gamma_1\!\cdot\!\vec\gamma_2-{\vec\gamma_1\!\cdot\!\vec
k\,\,\vec\gamma_2\!\cdot\!\vec k\over \vec k^2})\,].\label{3ee3}\end{equation}
This last gauge being not covariant, the space-time separation refers to the
laboratory frame. This gauge is not a priori a good choice for our covariant
calculation. In the two-fermion problem, however, it is well known that the
best choice for the calculation of the energy spectrum is Coulomb's gauge in
the center of mass reference frame. We could take this fact into account by
writing (\ref{3ee3}) for a pure two-fermion system in the two-fermion center of
mass reference frame $\,\vec P\!=\!0,\,$ making everything covariant by using
$\,P:$
\begin{equation}\beta_1\beta_2\to{(\gamma_1\!\cdot\! P)(\gamma_2\!\cdot\!
P)\over P^2},\qquad\vec a\!\cdot\!\vec b\to{(a\!\cdot\! P)(b\!\cdot\! P)\over
P^2}-(a\!\cdot\!b)\,,\qquad \label{3eeb3}\end{equation} and importing the
result into the two-fermion in an external potential problem, where $\,\vec
P\,$ is no more conserved. We could call this the "covariant Coulomb gauge". 
\par We have now to compute $\,V_e\,$ by using (\ref{p4}), with
$\,K^T_e\!=\!K\,$ given by (\ref{3ee2}). The principal difficulty comes from the
fact that the modified Dirac operators
$\,h_{ie}\,$ are diagonal neither in momentum space, nor in configuration
space. If we expand the potential $\,V_e\,$ into  the eigenfunctions of the
$\,h_{ie},\,$ it will be given by (in Feynman's gauge):
\begin{equation}V_e(\vec p_i^{\,'},\vec
p_i\,)\,=\,\sum_{{\omega'}_i\,\omega_i}\,\psi_{\omega'_1}(\vec
p_1^{\,'})\psi_{\omega'_2}(\vec p_2^{\,'})\,<\omega'_i\,\vert\,
V_e\,\vert\,\omega_i>\,\psi^+_{\omega_1}(\vec p_1)\psi^+_{\omega_2}(\vec
p_2)\label{3ee4}\end{equation} with
$$<\omega'_i\,\vert\,
V_e\,\vert\,\omega_i>\,=\,-\,{e^2\over2\pi^2}\,\hat\theta_e(\omega'_i)
\int\!d^3p'_1
d^3p'_2d^3p_1d^3p_2\,\psi^+_{\omega_1}(\vec p_1^{\,'})\psi^+_{\omega_2}(\vec
p_2^{\,'})$$
\begin{equation}{1-\vec\alpha_1\!\cdot\!\vec\alpha_2\over[\,\mu_e(\omega'_i)
-\mu_e(\omega_i)\,]^2
-(\,\vec p^{\,'}-\vec p\,)^2}\,\psi_{\omega_1}(\vec p_1)\psi_{\omega_2}(\vec
p_2)\,\hat\theta_e(\omega_i)\,\label{3ee5}\end{equation} where
$\,\psi_{\omega_i}(\vec p_i)\,$ is an eigenstate of $\,h_{ie}\,$ with
eigenvalue $\,\omega_i.\,$ The sum in (\ref{3ee4}) includes also an integral on
the continuum. This expression is well adapted to a perturbation calculation
starting with an eigenstate of $\,S_e.\,$ The first-order energy shift will
indeed be $\,<\omega_i\,\vert\, V_e\,\vert\,\omega_i>,\,$ in which
$\,\hat\theta_e(\omega_i)\,$ will be equal to 1, as
$\,P_0\!=\!\omega_1\!+\!\omega_2.$

\subsubsection{Positronium in an external potential.}  For the positronium in
an external potential, the mutual interaction kernel and potential in the Born
approximation (neglecting the annihilation graph) are the same as for the
two-electron pair above, with a change of sign. The external potentials could
again be the potentials generated by a nucleus, and we would write again
(\ref{3ee1}), with a change of sign for the positron. We could also consider
non-central potentials (as
$\,V_1\!=\!ex_{13},\,V_2\!=\!-ex_{23}\,$ for a constant electric field along
the third axis). The representation of $\,V_e\,$ in the space of the
eigenfunctions of the modified Dirac operators gives a meaning to our rather
formal expression (\ref{p4}), but it is not well adapted to this new framework,
as the principal interaction is now the binding mutual interaction. We shall
therefore isolate a zero-order Coulomb contribution from
$\,V_e:\,$ 
\begin{equation}V_e(\vec p_i^{\,'},\vec
p_i\,)=V^C(\vec p^{\,'},\vec
p\,)\,+\!\sum_{{\omega'}_i\,\omega_i}\,\psi_{\omega'_1}(\vec
p_1^{\,'})\psi_{\omega'_2}(\vec p_2^{\,'})<\omega'_i\,\vert\,
(V_e\!-\!V^C)\,\vert\,\omega_i>\psi^+_{\omega_1}(\vec
p_1)\psi^+_{\omega_2}(\vec p_2)\label{3e6}\end{equation} with
$$<\omega'_i\,\vert\,
(V_e\!-\!V^C)\,\vert\,\omega_i>\,=\,{e^2\over2\pi^2}\,\hat\theta_e(\omega'_i)
\int\!d^3p'_1
d^3p'_2d^3p_1d^3p_2\,\psi^+_{\omega_1}(\vec p_1^{\,'})\psi^+_{\omega_2}(\vec
p_2^{\,'})$$
\begin{equation}\left[\,{1-\vec\alpha_1\!\cdot\!\vec\alpha_2\over[\,
\mu_e(\omega'_i)-\mu_e(\omega_i)\,]^2
-(\,\vec p^{\,\,'}-\vec p\,)^2}\,+\, {1\over(\,\vec p^{\,\,'}-\vec
p\,)^2}\right]\,\psi_{\omega_1}(\vec p_1)\psi_{\omega_2}(\vec
p_2)\,\hat\theta_e(\omega_i)\,\label{3e7}\end{equation}             
 
\section{Three fermions.}

\subsection{3D equation.} It is easy to build a three-fermion 3D equation with
three two-fermion potentials $\,U_{ij}(P_{ij0})\,$ (we explicitate for a while
the dependence in the total energy of the two fermions):    
\begin{equation}(P_0-S)\,\psi\,=\,[\,U_{12}(P_0\!-\!h_3)\,+\,U_{23}
(P_0\!-\!h_1)\,+\,U_{31}(P_0\!-\!h_2)\,]\,\psi\label{42}\end{equation}  with
$\,S\!=\!h_1+h_2+h_3.\,$ We have replaced the arguments 
 $\,P_{ij0},\,$ by the operators
$\,P_0\!-\!h_k\,$ in order to get the exact cluster-separated limits. At the
$\,V_{23}=V_{31}=0\,$ limit, for example, we get indeed
\begin{equation}\bigg[\,P_0-h_1-h_2-h_3\,\bigg]\,\psi\,=\,
U_{12}(P_0\!-\!h_3)\,\psi.\label{43}\end{equation} Writing
\begin{equation}\psi=\psi_{12}\,\psi_3,\qquad
P_0\,=\,P_{120}\,+\,p_{30}\label{44}\end{equation} we get two independent
equations:
\begin{equation}\bigg[\,P_{120}-h_1-h_2\,\bigg]\,\psi_{12}\,=\,
U_{12}(P_{120})\,\psi_{12},\qquad
p_{30}\,\psi_3\,=\,h_3\,\psi_3\label{43.1}\end{equation} as
$\,P_0\!-\!h_3\,$ can be replaced by $\,P_0\!-\!p_{30}\!=\!P_{120}.\,$   Our 3D
equation (\ref{42}) satisfies thus clearly the cluster separability
requirement. Furthermore, the three cluster-separated limits are exactly the 3D
equations we would get by 3D-reducing the corresponding two-fermion
Bethe-Salpeter equations. This cluster separability is a property of the
equation, or, if we prefer, of the full Green function. For a given scattering
solution it is also possible to take the cluster-separated limit at fixed
$\,P.\,$ This is not possible for the bound state solutions. 
\par Our use of a covariant approached propagator results in an explicit
covariance of the cluster-separated limit equations like (\ref{43.1}): this
equation can directly be written in covariant form and each term of the series
giving the 3D potential is separately covariant. We could thus neglect all
terms of $\,K^T\,$ but the Born term without losing the covariance of the
cluster-separated limits.\par Let us recapitulate the elements of our 3D
equation (\ref{42}) (we give only $\,U_{12}):\,$
\begin{equation}U_{12}\,=\,V_{12}-\{P_0\!-\!S,\,{V_{12}\over2(P_0\!-\!h_3)}\,\}
+{V_{12}\,(P_0+S-2h_3)\,V_{12}\over4(P_0-h_3)^2}\label{3f5}\end{equation}
\begin{equation}V_{12}\,=\,-2i\pi\,\hat\theta_{12}\,\beta_1\beta_2\,
K^T_{12}(\mu_{12},\mu_{12})\,\hat\theta_{12}\label{3f6}\end{equation}
\begin{equation}\hat\theta_{12}\,=\,\theta(P_0-h_3-\vert\vec p_1\vert-
\sqrt{\vec p_2^{\,2}+m_2^2-m_1^2})\label{3f7}\end{equation}
\begin{equation}\mu_{12}\,=\,{1\over2(P_0-h_3)}\,(E_1^2-E_2^2)
\label{3f8}\end{equation}
\begin{equation}K^T_{12}\,=\,K_{12}+K_{12}G^R_{12}K_{12}+\cdots
\label{3f9}\end{equation}
$$G^R_{12}\,=\,\left({1\over(p_{10}-h_1+i\epsilon h_1)\,(p_{20}-h_2+i\epsilon
h_2)}\right.$$
\begin{equation}\left.+\,\,{2i\pi\over
P_0-S+i\epsilon}\,{P_0+S-2h_3\over2(P_0-h_3)}\,\delta(p_{120}-\mu_{12})
\,\hat\theta\right)
\,\beta_1\beta_2\label{3f10}\end{equation} We could keep only the Born terms in
the $\,K^T_{ij}\,$ without losing the covariance of the cluster-separated limit
equations.

\subsection{Two-body limits.} It is interesting to compare the two-body limits
(say $\,m_3\to\infty\,$) of our three-fermion equations with our two-fermion in
an external potential equations. Writing
$\,P_0\!=\!m_3+P_{120}\,$ and making $\,m_3\to\infty,\,$ we get
\begin{equation}P_{120}\,\psi\,=\,(\,h_1+h_2+V_1+V_2+U_{12}\,)\,\psi
\label{tb1}\end{equation}
where $\,U_{12}\,$ is still given by (\ref{3f5}), with $\,P_0\!-\!h_3\,$
replaced by $\,P_{120},\,$ while
\begin{equation}V_1\,=\,\theta\,(P_{120}-h_2-\vert\,\vec
p_1\vert\,)\,V_1^C\,\left(1-{\theta\,(-P_{120}+h_2+\vert\,\vec p_1\vert\,)\over
P_{120}-h_1-h_2}\,V_1^C\right)^{-1}\!\theta\,(P_{120}-h_2-\vert\,\vec
p_1\vert\,)\label{tb2}\end{equation}
\begin{equation}V_2\,=\,\theta\,(P_{120}-h_1-\vert\,\vec
p_2\vert\,)\,V_2^C\,\left(1-{\theta\,(-P_{120}+h_1+\vert\,\vec p_2\vert\,)\over
P_{120}-h_1-h_2}\,V_2^C\right)^{-1}\!\theta\,(P_{120}-h_1-\vert\,\vec
p_2\vert\,)\label{tb3}\end{equation} We note the following differences with the
two-fermion in an external potential equation:\par -- In the mutual interaction
term (which takes its symmetrized form), the free Dirac operators $\,h_i\,$ are
not replaced by the operators $\,h_{ie}.$\par -- The external potentials are
projected, as in the one-body limit (\ref{4.9}). The energy argument $\,W_1\,$
is here $\,P_{120}\!-\!h_2\,$ (and vice-versa). If $\,V_1^C\,$ is
energy-dependent (to consider more general cases than a pure Coulomb potential)
its energy argument will also be replaced by $\,P_{120}\!-\!h_2.\,$\par The
difference between both approaches consists clearly in three-body terms. The
cluster-separated limits are still exact, as they were before taking the
$\,m_3\to\infty\,$ limit. These cluster-separated limits appear when two of the
three two-body interactions are "switched off". When only the mutual
interaction $\,V_{12}\,$ is "switched off", we do not completely get two
independent equations as in section 3. Writing
$\,P_{120}=p_{10}+p_{20},\,$ we see that $\,V_1\,$ will depend on
$\,p_{20}\!-\!h_2\,$ and $\,V_2\,$ on $\,p_{10}\!-\!h_1.\,$ It must however be
noted that the external potentials obtained by the $\,m_3\to\infty\,$ limit are
less general than the external potential we could consider in the formalism of
section 3: here we expect central potentials decreasing with the distance. With
such potentials, it is not possible to let fermion 1 (for example) go to
infinity without having both $\,V_{12}\,$ and $\,V_{31}\,$ going to
zero.

\subsection{Examples.}
\subsubsection{Two electrons and a nucleus.} For two electrons and a nucleus of
spin ${1\over2},\,$ we have the Born terms
\begin{equation}K_{12}(p'_{12},p_{12})\,=\,{-2ie^2\over(2\pi)^3}\,{1\over
k_{12}^2+i\epsilon}\,(\gamma_1\!\cdot\!\gamma_2)\label{3e1}\end{equation}
\begin{equation}K_{23}(p'_{23},p_{23})\,=\,{2i\,Ze^2\over(2\pi)^3}\,{1\over
k_{23}^2+i\epsilon}\,(\gamma_2\!\cdot\!\gamma_3)\label{3e2}\end{equation}   
\begin{equation}K_{31}(p'_{31},p_{31})\,=\,{2i\,Ze^2\over(2\pi)^3}\,{1\over
k_{31}^2+i\epsilon}\,(\gamma_3\!\cdot\!\gamma_1)\label{3e3}\end{equation}
The 3D potentials are, in the momentum representation (we omit the momentum
arguments of the $\,V_{ij}\,$  here and in the next subsection):
\begin{equation}V_{12}={-e^2\over2\pi^2}\,\theta\,(P_0-h_3-\vert\,\vec
p^{\,'}_1\vert-\vert\,\vec
p^{\,'}_2\vert\,)\,{1-\vec\alpha_1\!\cdot\!\vec\alpha_2\over k^2_{120}-\vec
k_{12}^2}\,\theta\,(P_0-h_3-\vert\,\vec p_1\vert-\vert\,\vec
p_2\vert\,)\label{3e4}\end{equation}
$$V_{23}={Ze^2\over2\pi^2}\,\theta\,(P_0-h_1-\vert\,\vec
p^{\,'}_2\vert-\sqrt{\vec p^{\,\,'2}_3+m_3^2-m_2^2}\,)$$
\begin{equation}\,{1-\vec\alpha_2\!\cdot\!\vec\alpha_3\over k^2_{230}-\vec
k_{23}^2}\,\theta\,(P_0-h_1-\vert\,\vec p_2\vert-\sqrt{\vec
p_3^{\,2}+m_3^2-m_2^2}\,)\label{3e5}\end{equation} 
where $\,k^2_{120}\,$ and $\,k^2_{230}\,$
must be taken at 
\begin{equation}k^2_{120}\,=\,\left({\vec P_{12}\cdot\vec k_{12}\over
P_0-h_3}\right)^2,\qquad k^2_{230}\,=\,\left({\vec P_{23}\cdot\vec k_{23}\over
P_0-h_1}\right)^2,\label{pr2}\end{equation} while
$\,V_{31}\!=\!V_{23}\,(1\!\leftrightarrow\!2).\,$ We could also use the
"invariant Coulomb gauge" (in fact, one gauge for each two-fermion subsystem).
If we had started with a true three-fermion Bethe-Salpeter equation we could
not use three different gauges simultaneously, but here our aim is only to get
continuum dissolution-free equations with the correct cluster-separated limits.

\subsubsection{Three protons exchanging mesons.} For three protons exchanging
a vector, a scalar and a pseudoscalar meson, the contribution of the Born terms
to $\,V_{12}\,$ is
$$V_{12}={-1\over2\pi^2}\,\theta\,(P_0-h_3-\vert\,\vec
p^{\,'}_1\vert-\vert\,\vec p^{\,'}_2\vert\,)\,\left\{\,{g^2_{\rho}\over
k_{120}^2-\vec k_{12}^2-m^2_{\rho}}\,(
1-\vec\alpha_1\!\cdot\!\vec\alpha_2-{k_{120}^2-\vec\alpha_1\!\cdot\!\vec
k\,\,\vec\alpha_2\!\cdot\!
\vec k\over m^2_{\rho}}\,)\right.$$
\begin{equation}+\,\left.{g^2_{\sigma}\,\beta_1\beta_2\over k_{120}^2-\vec
k_{12}^2-m^2_{\sigma}}\,+\,{g^2_{\pi}\beta_1\beta_2\gamma_{15}\gamma_{25}\over
k_{120}^2-\vec k_{12}^2-m^2_{\pi}}\,\right\}\,\theta\,(P_0-h_3-\vert\,\vec
p_1\vert-\vert\,\vec p_2\vert\,)\label{pr1},\end{equation}  where where
$\,k^2_{120}\,$ must again be replaced by (\ref{pr2}), and similarly for
$\,V_{23}\,$ and
$\,V_{31}.$

\section{Conclusions.} Using the explicitly covariant Sazdjian propagator
combined with our covariant substitute of the positive-energy projector, we
built a 3D equation for two fermions in an external potential and for three
fermions. The two-fermion equations obtained at the cluster separated limits
are these which would be obtained by a 3D reduction of the two-fermion
Bethe-Salpeter equation. In our preceding works this implied that these
cluster-separated limits were implicitly covariant if not truncated, even if they
remained written in a specific reference frame (the laboratory frame or the
three-fermion rest frame). In the present work, however, they are explicitly
covariant, which implies that the covariance will survive the possible
truncations of the 3D potentials. The invariant substitute of the positive energy
projector we use brings a cutoff on the spatial momenta, which could be welcome
when insuring the existence of some integrals. This cutoff appear for rather low,
but nevertheless truly relativistic values. It is not by itself an approximation:
it reflects the choice to work with a given projection of the two-fermion
equation.\par The 3D equations we write in this work are more complicated that
the corresponding equations in our preceding works. Keeping in mind that in both
cases the 3D potentials are given by series to be truncated anyway, we hope that
the Lorentz invariance / cluster separability properties of the truncations of
the present equations will lead to a better approach of the real physics in the
first terms.

\end{document}